\documentclass[preprints,article,accept,moreauthors,pdftex,10pt,a4paper]{mdpi} 

\firstpage{1} 
\pubvolume{xx}
\issuenum{1}
\articlenumber{1}
\pubyear{2018}
\copyrightyear{2018}
\history{}

\usepackage{mycommands}
\usepackage[normalem]{ulem}  

\graphicspath{ {figs/} }  

\newcommand{\wt}[1]{\widetilde{#1}}

\newcommand{\Var}{\operatorname{Var}}

\renewcommand{\eref}[1]{Eq.~\eqref{#1}}
\Title{Investigation of a Multiple-Timescale Turbulence-Transport Coupling Method in the Presence of Random Fluctuations}

\Author{Jeffrey B.\ Parker*\orcidA{}, Lynda L.\ LoDestro, and Alejandro Campos}

\AuthorNames{Jeffrey B. Parker, Lynda L. LoDestro, Alejandro Campos}

\address[1]{%
\quad Lawrence Livermore National Laboratory, 7000 East Avenue, Livermore, CA 94550, USA}

\corres{Correspondence: parker68@llnl.gov}

\abstract{One route to improved predictive modeling of magnetically confined fusion reactors is to couple transport solvers with direct numerical simulations (DNS) of turbulence, rather than with surrogate models.  An additional challenge presented by coupling directly with DNS is that the inherent fluctuations in the turbulence, which limit the convergence achievable in the transport solver.  In this article, we investigate the performance of one numerical coupling method in the presence of turbulent fluctuations.  To test a particular numerical coupling method for the transport solver, we use an autoregressive-moving-average model to efficiently generate stochastic fluctuations with statistical properties resembling those of a gyrokinetic simulation.  These fluctuations are then added to a simple, solvable problem, and we examine the behavior of the coupling method.  We find that monitoring the residual as a proxy for the error can be misleading.  From a pragmatic point of view, this study aids us in the full problem of transport coupled to DNS by predicting the amount of averaging required to reduce the fluctuation error and obtain a specific level of accuracy.}

\keyword{multiple-timescale; transport; turbulence}

\begin{document}

\section{Introduction}
One challenge of magnetically confined fusion reactors is maintaining the plasma core at the high temperatures required for fusion to occur.  The temperature gradient between the hot  central plasma and the cooler edge is expected to be unstable, resulting in turbulent motions that carry heat toward the outside of the plasma.  The ability to accurately model the turbulence and its effect on plasma reactors in a predictive way would accelerate the progress of fusion research.  One challenge with quantitatively predicting the overall behavior is the enormous complexity of plasma turbulence and the large span of spatio-temporal scales in this process.  These scales are often categorized as turbulence and the slower transport.

Modeling the effect of turbulence on transport often occurs through transport equations, based on a scale separation.  `Transport solvers' numerically solve the transport equations, which are essentially local conservation equations, for macroscopic plasma profiles such as density and temperature.  These transport equations are coarse-grained such that the macroscopic profiles have slow variation in time and space compared to the turbulent fluctuations that are the source of the transport.  On the slow transport timescale and long length scales, the macroscopic profiles are constant on magnetic flux surfaces.  Hence, transport equations for these profiles are one-dimensional (1D) in space.  The transport equations will include terms, such as the flux of heat across a magnetic surface, which involve quantities arising due to turbulence, but that are presumed to be smoothly varying after averaging over the fast timescales and short length scales.  Other effects besides turbulent transport can be included in the transport equations, such as neoclassical transport or plasma heating or fueling.  The transport equation can be used to predict the spatiotemporal evolution of the temperature profile and other quantities (for example, \cite{kim:2016}).  The decomposition into turbulence and transport should be valid as long as there is a strong scale separation, which is believed to often be the case in the core of a tokamak.

\begin{figure}
	\centering
	\includegraphics[width=4in]{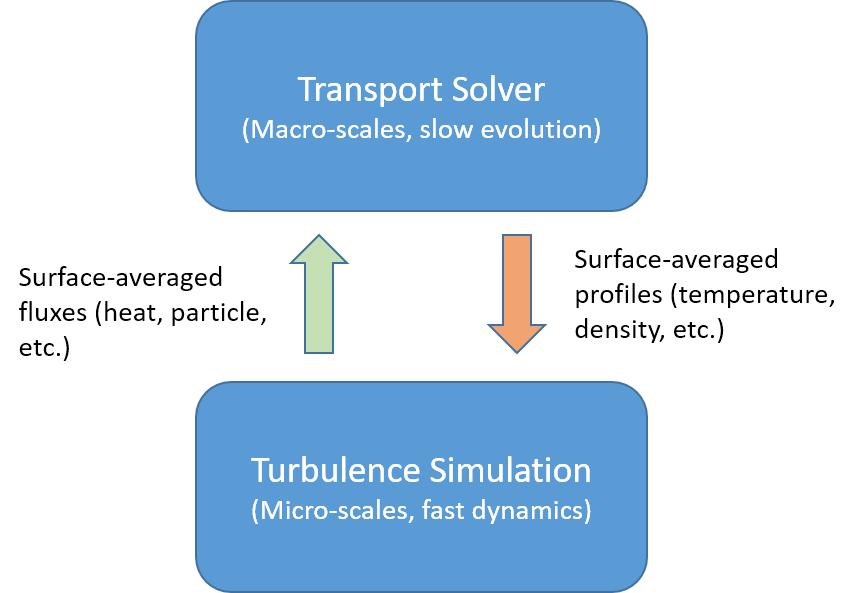}
	\caption{Schematic diagram of coupling between transport solver and turbulence simulation for a self-consistent solution.}
	\label{fig:schematic}
\end{figure}

Presently, a common approach for modeling the effect of turbulence on transport is to calculate the turbulent fluxes using a \emph{surrogate model} in lieu of directly simulating the turbulence itself.  The use of surrogate models allows modeling plasma transport in a computationally reasonable manner.  Entire tokamak discharges can be modeled, from start-up to flat-top to shut-down.  Of course, surrogate models exist on a spectrum with more and less elaborate models.  Very simple analytic models are used sometimes \cite{tang:1986,jardin:1993,erba:1997}.  A method closer to the fundamental physics, which is in widespread use now, is that of quasilinear transport models \cite{kinsey:2011,bourdelle:2007}.  This approach assumes certain phase relationships based on linear eigenmodes, along with estimates of the turbulent amplitudes, to predict a turbulent flux.  Even this approach can be time consuming, and so one method of providing a faster surrogate model to the quasilinear transport models is through the fitting of neural networks \cite{citrin:2015,meneghini:2017}.  The tradeoff for the speed of surrogate models is their uncertain accuracy, particularly when trying to predict new physics or regimes.

As computers become more powerful, another option is becoming feasible: coupling direct numerical simulations (DNS) of turbulence directly to the transport equations.  Rather than invoking a simplified formula to capture what the turbulent heat flux should be, one can instead simulate it.  Turbulence simulations are more computationally demanding than the surrogate models by many orders of magnitude.  Yet with increasing computational resources, the promise of more realistic modeling may be attractive enough to bear the costs.  A high-level schematic diagram of how the computational coupling occurs is shown in Figure~\ref{fig:schematic}.  A few attempts have been made along this path aimed at plasma turbulence simulations, including some attempts with gyrokinetic simulations \cite{shestakov:2003,candy:2009,barnes:2010,parker:2018,highcock:2018}.  However, the problem is far from solved and many challenges remain for this ambitious program.
 

A consideration that arises when coupling a transport solver to a turbulence simulation is how to confront the inherent turbulent fluctuations.  Because of the timescale separation between the slow timescale and the turbulence, the quantities of interest relevant to the slow timescale involve only the time average of turbulent quantities.  A concrete example is the heat flux.  In turbulence simulations, the surface-averaged heat flux is a primary quantity of interest.  To evolve the pressure profile on the slow timescale, a transport solver needs only the time average of this flux.  From that perspective, fluctuations of the flux about its true mean value are undesirable, as they introduce an error.  The effect of these fluctuations and the error they cause in the solution of the transport equations must be understood and managed.  One cannot allow the effect of the fluctuations to be so large that the error in the solution is unacceptably large, yet one does not want to spend more effort than necessary minimizing the effect of the fluctuations, which almost invariably requires additional computational expense.  Understanding such tradeoffs is the first step in determining how to achieve a manageable balance between a reliable transport solver and overall computational cost.

The purpose of this paper is to study the behavior of a transport solver in the presence of fluctuations in the surface-averaged flux.  The considerations we introduce apply equally to any algorithm used in a transport solver with implicit timesteps, although we will focus our numerical investigation on the method described in Refs.~\cite{shestakov:2003,parker:2018}.

It should be noted that the entire issue of fluctuations does not exist in the surrogate methods previously described, such as quasilinear models or machine-learning-based fits.  Those models for calculating turbulent flux have the property that for a given input (typically macroscopic profiles),  they always yield the same output.  In contrast, a turbulence simulation yields different values of the heat flux when sampled at different times, even in a statistically steady state.  The problem of dealing with fluctuations in the transport solver arises only when coupling to actual turbulence simulations.

We adopt the mindset that from the point of view of the slow timescale, a statistical framework is useful for analyzing the turbulent fluctuations.  To investigate and characterize the behavior of the transport solver, a conceptual simplification is to consider the turbulent fluctuations as \emph{random} signals about their true mean.  True randomness is merely a convenient assumption, because the gyrokinetic equations describing the turbulence are nonrandom and deterministic; however, turbulent fluctuations are typically chaotic with sensitive dependence on initial conditions.

We study the behavior of a transport solver in a simple problem with added random flctuations.  For systematic study, trying to generate many independent samples from the turbulence simulation would be extremely computationally costly.  Instead, we use an autoregressive-moving-average model to efficiently generate random signals.



The outline of this paper is as follows.  In Section \ref{sec:theorybackground}, we will describe the problem of interest, which is to solve a transport equation coupled with gyrokinetic turbulence simulations in a multiple-timescale framework.  Then, in Section \ref{sec:multiple_timescale_algorithm}, we will describe one numerical coupling method that has been used to solve this problem.  Section \ref{sec:math_considerations} introduces some basic mathematical considerations for characterizing fluctuations within a statistical framework.  In Section \ref{sec:iteration_analytic_model}, we analyze a simple analytic model to gain intuition about convergence in the presence of fluctuations.  In Section \ref{sec:generate_noise}, we describe a technique for efficiently generating fluctuations with certain desired statistical properties.  Then, in Section \ref{sec:numerical_behavior}, we investigate the behavior of the numerical coupling method in the presence of fluctuations.  Finally, we summarize our findings in Section \ref{sec:discussion}.


\section{Theoretical background: multiple-scale turbulence and transport framework}
\label{sec:theorybackground}
Our motivating model is the coupled set of gyrokinetic turbulence and transport equations appropriate for tokamaks \cite{sugama:1997,sugama:1998,abel:2013}.  For full details, the reader is referred to the literature; a simplified overview is given here.  The description of the fast, small-scale turbulence consists of a gyrokinetic equation for each species and Maxwell's equations.  The gyrokinetic equation evolves the gyrocenter distribution of each species in a 5D phase space (three space and two velocity dimensions).  The description of the slow, macroscopic transport evolution consists of local conservation equations for the surface-averaged density, momentum, and energy, which depend on only one spatial dimension, the flux coordinate.

However, for the purposes of this paper, we use a simplified paradigm model, which suffices to allow us to explore aspects of the numerical coupling method.  This simplified problem eliminates the toroidal geometry and focuses on a single transport equation.  We take a generic transport equation in a Cartesian geometry:
	\begin{equation}
		\pd{p}{t} + \pd{q}{x} = S,
		\label{paradigm_transport}
	\end{equation}
where $p(x,t)$ is a macroscopic profile, $q(x,t)$ is the surface-averaged flux of quantity $p$, and $S(x,t)$ includes all local sources.  The turbulent flux $q$ is to be computed from some independent model, and it is assumed that $q$ is a functional of the profile $p$, i.e., $q = q[p]$.  

In this paper, $q$ will be computed from an analytic model with added fluctuations rather than from DNS.


\section{A numerical method for multiple-timescale turbulence and transport}
\label{sec:multiple_timescale_algorithm}
We briefly review a numerical method for coupling a turbulence simulation and transport solver.  The method was introduced by Shestakov et al.\ for coupling transport to Hasegawa--Wakatani fluid turbulence \cite{shestakov:2003}, and was later used by Parker et al.\ for coupling transport to gyrokinetic turbulence in a tokamak \cite{parker:2018}.  Essentially, the method prescribes a procedure for iterating to convergence the nonlinear equation arising from an implicit timestep of \eref{paradigm_transport}, and in particular for handling the turbulent flux which depends nonlinearly on the profile $p$.

We introduce the subscript $m$ to denote the time index and $l$ to denote the iteration index within a timestep.  With a backward-Euler timestep, the iteration scheme is given by
	\begin{equation}
		\frac{p_{m,l} - p_{m-1}}{\D t} + \pd{}{x} \left( -D_{m, l-1} \pd{}{x} p_{m,l} + c_{m, l-1} p_{m,l} \right) = S_m,
		\label{iteration_scheme}
	\end{equation}
where $p_{m-1}$ is the converged (in $l$) profile of the previous timestep, and
	\begin{align}
		D_{m, l-1} &= -\th_{m,l-1} \ol{q}_{m, l-1} / \partial_x \ol{p}_{m, l-1} \ , \\
		c_{m, l-1} &= (1 - \th_{m,l-1}) \ol{q}_{m, l-1} / \ol{p}_{m, l-1} \ , \\
		\ol{q}_{m, l-1} &= \a q[\ol{p}_{m, l-1}] + (1 - \a) \ol{q}_{m, l-2} \ , \\
		\ol{p}_{m, l-1} &= \a p_{m, l-1} + (1 - \a) \ol{p}_{m, l - 2} \ . \label{iteration_scheme_pbar}
	\end{align}
Here, $\a$ is a relaxation parameter with $0 < \a \le 1$.  The notation $q[\ol{p}]$ indicates that the simulation that calculates $q$ is to be performed with the relaxed profile $\ol{p}$.  In \eref{iteration_scheme}, $q$ has been represented in terms of diffusive and convective contributions.  $D$ and $c$ are artificial transport coefficients, chosen in a way such that if the iteration converges, the sum from the diffusive and convective fluxes is exactly equal to $q$.  $\th$ is a parameter to control the diffusive-convective split and various strategies for choosing it are possible.  Within this paper, we choose $\th$ to be 1 or close to 1.

Space is discretized through second-order finite differences.  Equation~\eqref{iteration_scheme} becomes a matrix equation, schematically in the form
	\begin{equation}
		M_{l-1} p_l = g,
		\label{iteration_scheme_matrix_equation}
	\end{equation}
where $p_l$ is a vector, $g$ includes the source and $p_{m-1}$ terms, and $M$ is a matrix that depends on the values of the transport coefficients $D, c$.  The $m$ subscript has been suppressed here for simplicity.  We emphasize that $M$ in this equation is determined only by known, past iterates of $p$, as indicated by the subscript $l-1$.  This matrix equation can be solved for the new iterate by inverting, which amounts to solving a tridiagonal system.  This method has been implemented in the open-source Python code \texttt{Tango} \cite{tangocode}.

\section{Review of relevant mathematical considerations in a statistical framework}
\label{sec:math_considerations}
Inherent to turbulence are the chaotic, unpredictable fluctuations occurring on a fast timescale.  These fluctuations are a nuisance from the practical perspective of converging an implicit timestep in a transport solver.  This section reviews some statistical considerations relevant to a transport solver.  The concepts here underpin the rest of this article.

From the point of view of the macroscopic dynamics, the turbulent fluctuations can for some purposes be modeled as random because the timescale of the fluctuations is much shorter than the macroscopic timescale.  Henceforth, we will sometimes refer to the fluctuations as \emph{noise}.


The turbulent quantity of interest for the transport solver is the flux, such as the heat flux.  We will assume that the turbulence possesses a steady state where the statistical moments of all quantities are stationary.  Statistical moments of the heat flux, such as the mean, variance, skewness, and kurtosis, can be used to characterize the distribution of the flux.  Although the mean of the heat flux is small---changes to the macroscopic profile in our multiscale physics problem are slow by assumption---there is no restriction on the variance of the heat flux.  That is, the fluctuations in the heat flux about its mean value can be comparable to its mean value.  Indeed, large fluctuations are often observed in gyrokinetic simulations, as seen later in Figure~\ref{fig:turbulent_flux}(a).  Turbulent fluctuations, including the heat flux, are generated from nonlinear processes and so will not in general be normally distributed.

In the computational framework of coupling a turbulence simulation to a transport solver, the natural strategy is to pass the time average of the heat flux to the transport solver.  The heat flux, averaged over a time $T$, is given by
	\begin{equation}
		\avg{q}(x, t) = \frac{1}{T} \int_{t-T/2}^{t+T/2} dt'\, q(x, t').
		\label{flux_time_avg}
	\end{equation}
In discrete form, $\avg{q}$ is a sum over many realizations of the turbulent fluctuations.  If the averaging time is sufficiently long, one expects $\avg{q}$ to approach a normal distribution, even though $q$ at different times are not independent.

In statistical parlance, let $X_i$ be a stationary process with expected value $\operatorname{E}(X_i) = \m$ and variance $\Var(X_i) = \sigma^2$.  We do \emph{not} assume the $X_i$ are independent.  In addition to the statistical moments, another important statistical quantity is the autocorrelation time $\t$.  On an intuitive level, $\t$ is the time it takes for the process to ``forget'' its state.  Measured discretely in number of samples, the autocorrelation time is defined as
	\begin{equation}
		\t \defineas 1 + 2 \sum_{k=1}^\infty \r_k,
		\label{autocorrelation_definition}
	\end{equation}
where the autocovariance function $\gamma_k$ and autocorrelation function $\rho_k$ are defined as
	\begin{align}
		\gamma_k &\defineas \operatorname{Cov}(X_i, X_{i+k}), \\
		\rho_k &\defineas \frac{\gamma_k}{\gamma_0} = \frac{\gamma_k}{\sigma^2},
	\end{align}
and $\operatorname{Cov}(X,Y)$ is the covariance of $X$ and $Y$.

The autocorrelation time is important because it emerges naturally in the central limit theorem for correlated sequences \cite{billingsley:book}.  That theorem can be stated as follows: let $\ol{X}_n$ be the sample mean of $(X_1, \ldots, X_n)$ and suppose certain technical niceness conditions on the $X_i$ are satisfied.  Then, as $n \to \infty$,
	\begin{equation}
		\frac{\ol{X}_n - \m}{\sigma / \sqrt{n_{\text{eff}}}} \Rightarrow N(0,1),
	\end{equation}
where $n_{\text{eff}} \defineas n/\tau$ and $N(a,b)$ is a normally distributed random variable with mean $a$ and variance $b$.  In other words, the distribution of $\ol{X}_n$ converges to $N(\m, \s^2 / n_\text{eff})$ as $n \to \infty$.

The theorem provides a theoretical basis for recognizing the importance of the parameter $n / \tau$ for determining the variance.  The number of samples is naturally measured in terms of the autocorrelation time.  Because the sequence is correlated, it takes approximately $\tau$ draws to obtain an independent sample, and so $n_\text{eff}$ is the number of effective independent samples.  Physically, $\tau$ is probably on the order of a turbulent eddy time.  Estimating $\t$ from data is discussed in the Appendix.

\section{Simple iteration problem with relaxation and noise}
\label{sec:iteration_analytic_model}
We consider a simple mathematical model that contains some of the key features of the transport iteration described in Section~\ref{sec:multiple_timescale_algorithm}, including the feature of fixed-point iteration plus relaxation.  The model will illustrate some of the expected behavior and scalings in the presence of random noise.

The classic fixed-point iteration is
	\begin{equation}
		x_{n+1} = f(x_n),
	\end{equation}
which has solutions $x_*$ satisfying $x_* = f(x_*)$.  Here, we consider a variation on this iteration, given by
	\begin{align}
		\wt{f}_n &= f(x_n) + \ve_n, \label{simple_iteration:first} \\
		\ol{f}_n &= \a \wt{f}_n + (1 - \a) \ol{f}_{n-1}, \label{simple_iteration:second} \\
		x_{n+1} &= \ol{f}_n. \label{simple_iteration:third}
	\end{align}
This system is reminiscent of Eqs.~\eqref{iteration_scheme}--\eqref{iteration_scheme_pbar} with $x \sim p$, $\wt{f} \sim q$, and $\ol{f} \sim \ol{q}$.  We again seek solutions satisfying $x_* = f(x_*)$.  Equation \eqref{simple_iteration:first} states that at each iteration, a random component $\ve_n$ is added to the basic functional response $f(x_n)$ to give $\wt{f}_n$.  We motivate \eref{simple_iteration:first} in the following way.  Consider the quantity $\wt{f}(x, t)$ to be like the turbulent flux, in that it depends on the input $x$ and fluctuates in time.  In a statistically steady state, let $\avg{\wt{f}(x, t)}_t \defineas f(x)$ denote the time average of $\wt{f}(x, t)$, and let $v(x, t) \defineas \wt{f}(x, t) - f(x)$.  Then $\wt{f} = f(x) + v(x, t)$.  For simplicity, we assume that the statistical properties of $v(x, t)$ do not depend on $x$.  We further collapse the time dependence to a random variable, $v(t) \to \ve_n$, yielding \eref{simple_iteration:first}.  We take $\ve_n$ to be a centered normal random variable, $\ve_n \sim N(0, \s^2)$, with each $\ve_n$ independent.  The normality of $\ve_n$ may be justified in some situations by invoking the central limit theorem.  Equation \eqref{simple_iteration:second} represents a relaxation and \eref{simple_iteration:third} represents the functional iteration using the relaxed $\ol{f}$.

We define the error $\de_n$ and the residual $\eta_n$:
	\begin{align}
		\de_n &\defineas x_n - x_*, \label{simple_iteration:error} \\
		\eta_n &\defineas \ol{f}_n - x_n = x_{n+1} - x_n. \label{simple_iteration:residual}
	\end{align}
	
\begin{figure}
	\centering
	\includegraphics{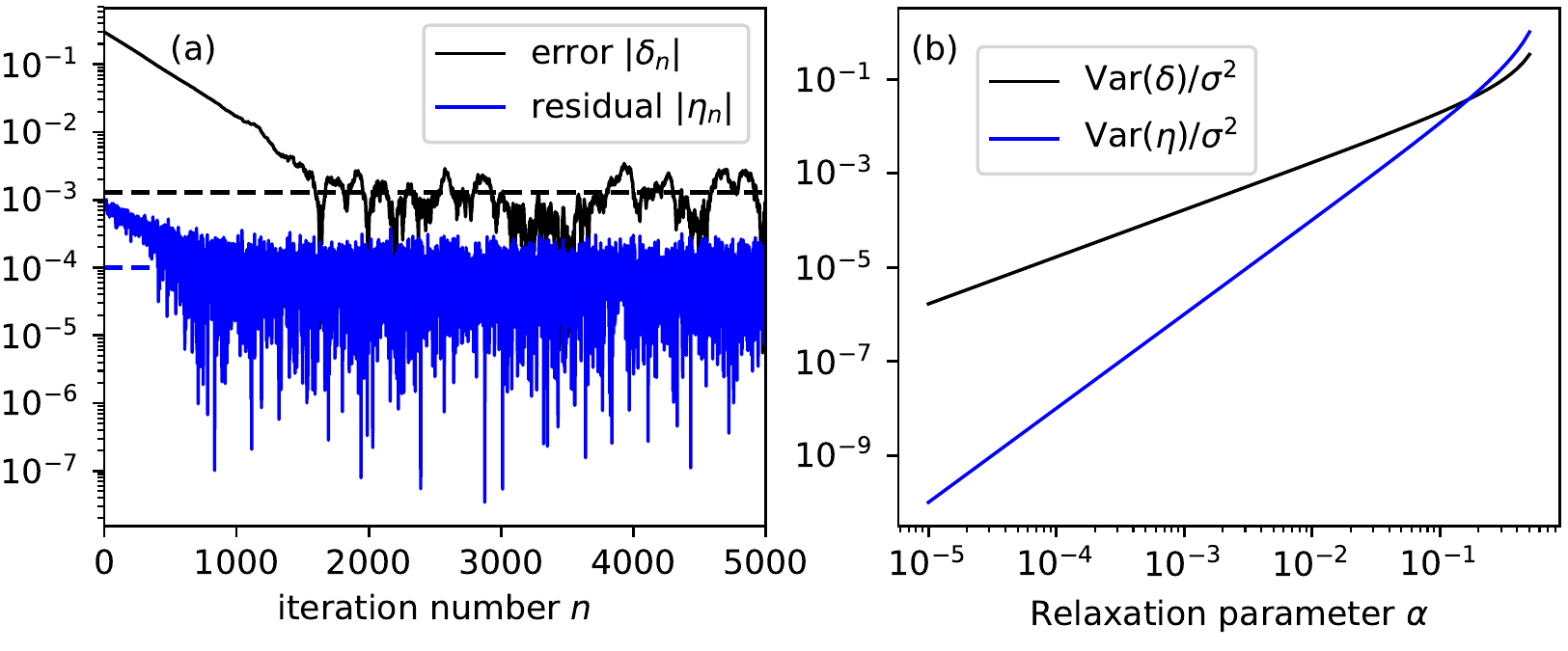}
	\caption{(\textbf{a}) Plots of the error $|\de_n|$ and residual $|\eta_n|$ as a function of iteration number for one realization of the system in Eqs.~\eqref{simple_iteration:first}--\eqref{simple_iteration:third}.  For this example, we take $f(x) = 2 - x^2$, $x_0 = 0.7$, $\a = 0.001$, and $\s=0.1$.  The horizontal dashed lines are the theoretical standard deviations of $\de$ and $\eta$ as $n \to \infty$.  Observe that as $x_n \to x_* = 1$, the residual reaches its noise floor at an earlier iteration number ($n \approx 700$) than the actual error reaches its noise floor ($n \approx 1700$).  (\textbf{b}) The scaling of $\Var(\de)$ and $\Var(\eta)$ as the relaxation parameter $\a$ varies.  In the small-$\a$ regime, $\Var(\de) \sim \a$ and $\Var(\eta) \sim \a^2$.}
	\label{fig:eta_delta}
\end{figure}

As an example, we take $f(x) = 2 - x^2$.  In the absence of noise, $x_* = 1$ is a stable fixed point when $\a < 2/3$.  One realization using $\a = 0.001$ and $\s = 0.1$ is shown in Figure~\ref{fig:eta_delta}(a), which depicts $\de_n$ and $\eta_n$ as $x_n$ is iterated toward the solution $x_*=1$.  

Linearizing \eref{simple_iteration:error} about the solution $x_*$, we find the error behaves as
	\begin{equation}
		\de_{n+1} = \l \de_n + a_n,
		\label{simple_iteration:linearized_error}
	\end{equation}
where $\l \defineas 1 + \a (f'_* - 1)$, $f'_* \defineas f'(x_*)$, and $a_n \defineas \a \ve_n$.  We assume $|\l| < 1$.  Note that $\ve_n$ and $\de_n$ are independent.  We now examine the stationary limit $n \to \infty$, where statistics do not depend on $n$.  This limit defines the \emph{noise floor}, where the iteration process cannot converge closer to $x_*$ due to the random noise.  Taking the mean and variance of \eref{simple_iteration:linearized_error}, we find
	\begin{align}
		\lim_{n \to \infty} & \operatorname{E} (\de_n) = 0, \\
		\lim_{n \to \infty} & \Var (\de_n) \defineas \Var(\de) = \frac{ \a^2 \s^2}{1 - \l^2}.
	\end{align}	
Observe that $1 - \l^2 = 2\a(1 - f'_*) - \a^2 (1 - f'_*)^2$.  One may choose $\a$ small enough such that $\a ( 1-f'_*) \ll 1$.  We call this the ``small-$\a$ regime'' and use $1 - \l^2 \approx 2 \a (1 - f'_*)$.  In this regime,
	\begin{equation}
		\Var(\de) = \frac{\a \s^2}{2(1 - f'_*)}.
	\end{equation}
Hence, the standard deviation or root mean square (rms) of the error scales as $\a^{1/2} \s$ in the small $\a$ regime.

The behavior of the residual at the noise floor differs in its scaling.  Returning to \eref{simple_iteration:residual} and linearizing, we obtain $\eta_n = \a (f'_* - 1) \de_n + \a \ve_n$.  We take the variance of this expression, again using independence of $\de_n$ and $\ve_n$, and obtain
	\begin{equation}
		\Var (\eta_n) = \a^2 (1 - f'_*)^2 \Var(\de_n) + \a^2 \s^2.
		\label{simple_iteration:residual_variance}
	\end{equation}
So far, the noise floor limit has not been taken.  We take that now, and find
	\begin{equation}
		\lim_{n \to \infty} \Var(\eta_n) \defineas \Var(\eta) \approx \a^2 \s^2.
		\label{simple_iteration:residual_variance_noise_floor}
	\end{equation}
Here, the dominant contribution to $\Var(\eta)$ arises from noise \emph{at the current iterate}.  The contribution to $\Var(\eta)$ from $\de_n$ (which indirectly contains contributions from previous iterates of noise) is subdominant in the small-$\a$ regime.  An interesting consequence is that the rms residual at the noise floor scales as $\a \s$, which is \emph{different} than the $\a^{1/2} \s$ scaling of the rms error.  The different scalings are depicted in Figure~\ref{fig:eta_delta}(b).

Using the linearized equations, one can also solve for the $n$ dependence of $\de_n$ or $\eta_n$, rather than just the statistics in the $n \to \infty$ limit.  Until $\de_n$ is small, it decreases exponentially, following the convergence rate as if there were no noise.  But when $\de_n$ gets small enough and hits the noise floor, it fluctuates around zero with variance $\Var(\de)$, as depicted in Figure~\ref{fig:eta_delta}(a).

The residual $\eta_n$ will similarly decrease exponentially and hit a noise floor.  However,  the noise floor for the residual is encountered at an earlier iteration than when the noise floor for the error is encountered.  $\eta$ reaches its noise floor when, on the right-hand-side of \eref{simple_iteration:residual_variance}, the first term becomes smaller than the second term.  But at this iteration number, $\de_n$ is still decreasing.  This behavior can be seen in Figure~\ref{fig:eta_delta}(a).

One conclusion is that even when the residual has hit a noise floor and is not decreasing further, the iterates $x_n$ may still be getting better, measured by a reduction in $\de_n$.  However, one cannot evaluate $\de_n$ in practice without knowledge of $x_*$.  On the other hand, the residual is easy to compute, making it tempting to monitor the residual to determine convergence.  But as we have seen, the residual is not a perfect proxy for the solution error.  One possible way to improve the use of the residual as a proxy, which is not explored here, is to average the residual over several iterates, which could reduce the direct contribution of the random noise to a subdominant level.

\section{Generating fluctuations with certain properties for testing purposes}
\label{sec:generate_noise}
Investigating our turbulence-transport coupling method with added fluctuations requires the ability to efficiently generate time traces of fluxes with random noise, in a controlled fashion.  Such a capability facilitates more rapid exploration and testing than relying only on expensive turbulence simulations.  Here, we describe a method to generate random noise with certain properties, which can then be combined with some externally given signal that represents the mean value of the heat flux.

We will consider two possibilities.  First, we consider Gaussian noise with a finite spatial correlation length but white in time.  Second, we consider non-Gaussian noise that has finite correlation in time and is perfectly correlated in space.  

\subsection{Generating spatially correlated Gaussian noise}
\label{sec:generating_gaussian_noise}
To generate noise with a finite spatial correlation length, we use a discrete Langevin equation.  This model has the form
	\begin{equation}
		z_{n+1} = \lambda z_n + \ve_n,
		\label{discrete_langevin}
	\end{equation}
where $0 < \lambda < 1$, $\ve_n$ is a centered, normally distributed variable, and $n$ is a spatial grid index.  The correlation length (using the same definition as in \eref{autocorrelation_definition}), measured in discrete samples, is $(1+\lambda) / (1-\lambda)$.

\begin{figure}
	\centering
	\includegraphics{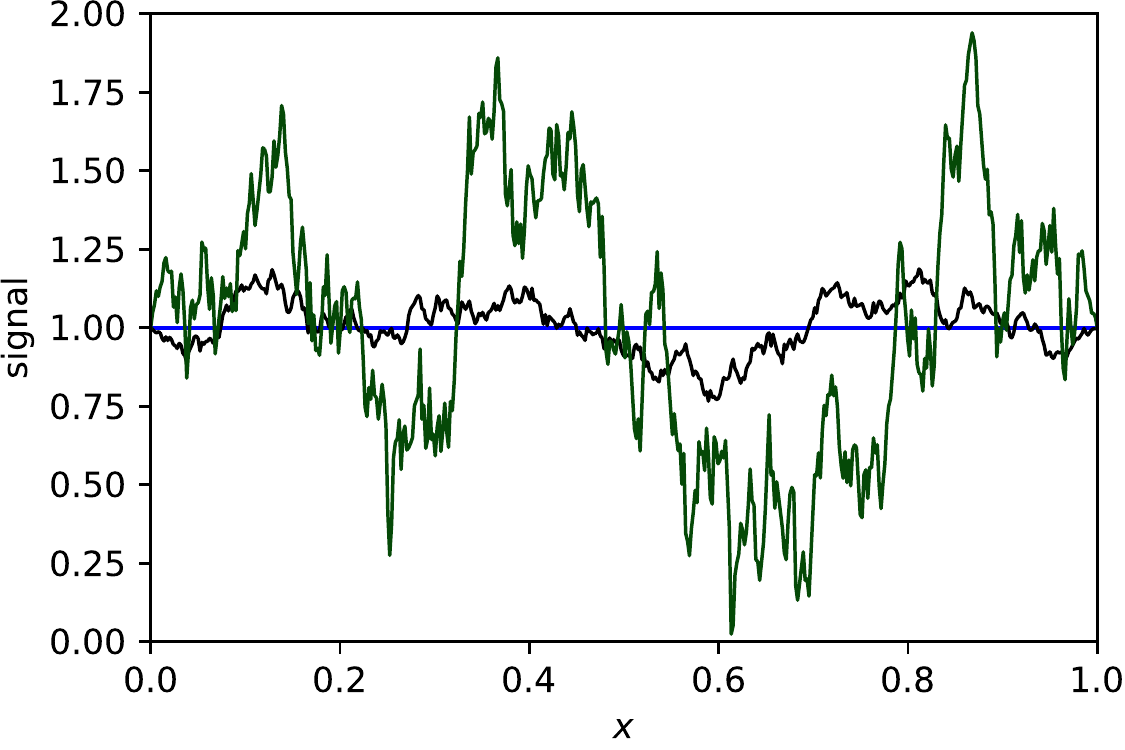}
	\caption{Realizations of Eqs.~\eqref{discrete_langevin} and \eqref{discrete_langevin_combined} for a spatially correlated process, with $s(x)=1$.  The correlation length is 0.2.  The black line has a standard deviation of 0.1 and the green line has a standard deviation of 0.5.}
	\label{fig:ar1_space_signal}
\end{figure}

In this way, at each timestep we construct a signal $z$ of length $N$, where $N$ is the number of spatial grid points, to be combined with an analytic signal.  If the analytic signal is $s(x)$, then our combined signal is
	\begin{equation}
		\wt{s}(x) = s(x) \bigl[ 1 + W(x) z(x) \bigr].
		\label{discrete_langevin_combined}
	\end{equation}
As a practical matter, to adhere to boundary conditions we have applied a tapering window $W(x)$ to $z$ that is equal to one in the interior and falls smoothly to zero at the boundaries.  The expected value of $\wt{s}(x)$ is $s(x)$.  Two example realizations are shown in Figure~\ref{fig:ar1_space_signal}.

\subsection{Generating temporally correlated non-Gaussian noise}
\label{sec:generating_non_gaussian_noise}
\begin{figure}
	\centering
	\includegraphics[width=\textwidth]{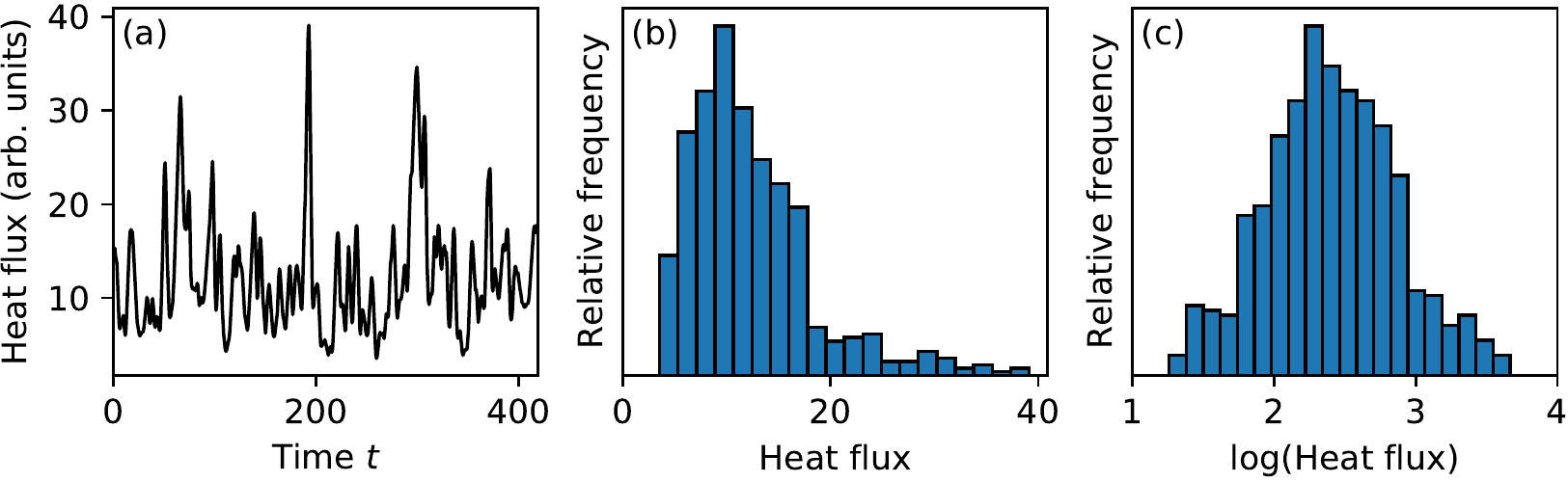}
	\caption{(\textbf{a}) Time trace of the ion heat flux (arbitrary units) from a typical gyrokinetic turbulence simulation with the GENE code.  Time is measured in $R_0 / v_{Ti}$.  The time trace was obtained in a statistically steady state.  The ion heat flux is averaged over a magnetic surface and averaged over a small radial window with width of several gyroradii.  In this time trace, the mean value is about 12.3, although bursts up to several times larger occur.  (\textbf{b}) Histogram of the heat flux values; the distribution is non-symmetric with a long tail.  (\textbf{c}) Same as (b), except the histogram is of the logarithm of heat flux values.  The distribution of the logarithm of the heat flux looks roughly symmetric and much closer to normally distributed than the distribution of the heat flux itself.}
	\label{fig:turbulent_flux}
\end{figure}

As we turn to the generation of fluctuations with a finite correlation in time, it is instructive to first examine an actual time trace of heat flux from a gyrokinetic simulation.  Clearly, it is desirable for artificial fluctuations to resemble the fluctuations occurring in practice.  Figure~\ref{fig:turbulent_flux}(a) shows a typical time trace of the ion heat flux from a gyrokinetic simulation; the heat flux has been surface-averaged and also averaged over a small radial window with width of several gyroradii.   This simulation was performed with the GENE code \cite{jenko:2000,gorler:2011,gorler:2011b,genecode}.  In the arbitrary units used in the figure, the heat flux has a mean value of about 12.3, although bursts up to several times larger than the mean can be observed.  The right-tailed nature of the distribution of the heat flux can be visualized in a histogram, as in Figure~\ref{fig:turbulent_flux}(b).  Using the batch means method described in the Appendix, we estimate the autocorrelation time to be $\t \approx 11.7$ samples.  The interval between samples is $\D t \approx 0.7$ $R_0 / v_{Ti}$, where $R_0$ is the major radius and $v_{Ti}$ is the ion thermal velocity.

For generating fluctuations correlated in time similar to those in Figure~\ref{fig:turbulent_flux}(a), we use the autoregressive-moving-average models ARMA($p,q$) from the field of time-series analysis \cite{box:book,wei:book}.  The ARMA($p,q$) model is written as
	\begin{equation}
		X_t = c+ \vp_1 X_{t-1} + \cdots + \vp_p X_{t-p} + \ve_t + \th_1 \ve_{t-1} + \cdots + \th_q \ve_{t-q},
	\end{equation}
where $X_j$ is the signal at time $j$, $c$ is the mean value, $p$ is the number of autoregressive (AR) terms, $q$ is the number of moving-average (MA) terms, $\vp_j$ are the AR parameters, $\th_j$ are the MA parameters, and the $\ve_j$ are independent and identically distributed centered Gaussian variables.  Equation~\eqref{discrete_langevin} is, in this framework, just the ARMA(1,0) or AR(1) model.

Given a finite sequence, there are established procedures to estimate optimal parameters $(p,q)$ as well as the parameters $\vp_j$ and $\th_j$ \cite{box:book, broersen:2002, broersen:2003}.  We use the ARMASA software package \cite{broersen:mathworks} for Matlab to perform the estimation on the heat flux in Figure~\ref{fig:turbulent_flux}(a).

\begin{figure}
	\centering
	\includegraphics{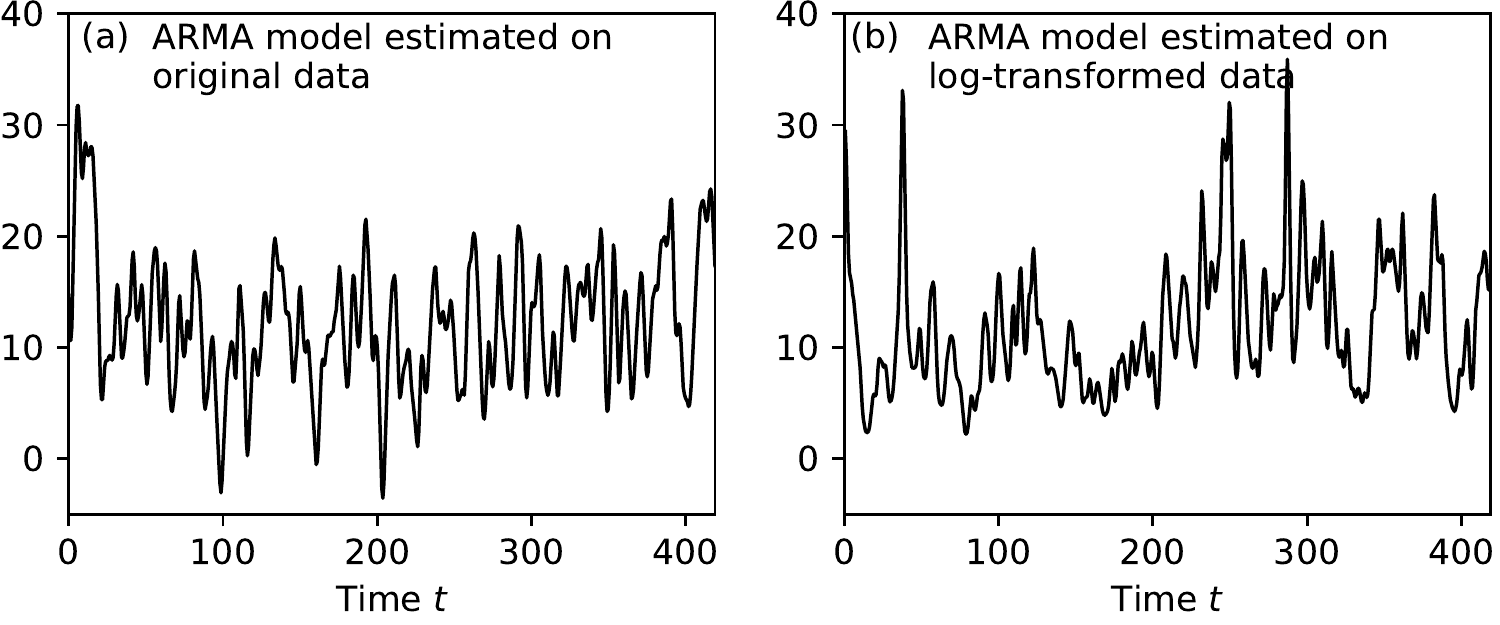}
	\caption{Generated signals using ARMA-estimated models of the heat flux in Figure~\ref{fig:turbulent_flux}(a).  (\textbf{a}) Realization of an ARMA model applied to the heat flux directly.  The output is Gaussian distributed and has negative values.  (\textbf{b}) Realization of an ARMA model applied to the logarithm of the heat flux, followed by reverse transforming as in \eref{arma:convexitycorrection}.  The output is log-normal distributed, with no negative values and with regular large bursts.}
	\label{fig:arma_signals}
\end{figure}

Fitting an ARMA($p,q$) model directly to the time trace in Figure~\ref{fig:turbulent_flux}(a) provides a less-than-ideal representation.  The reason is that ARMA models produce a Gaussian signal $X_t$, while the ion heat flux is non-Gaussian, as seen in Figure~\ref{fig:turbulent_flux}(b).  The ARMA parameters fit directly to this time trace are $p=3$, $q=2$, $\vp_1 = 2.2$, $\vp_2 = -1.7$, $\vp_3 = 0.51$, $\th_1 = 0.44$, and $\th_2 = -0.13$.  Using the method of estimating $\t$ directly from the ARMA parameters, described in the Appendix, gives $\tau \approx 14.7$, somewhat close to the estimate from batch means.  A realization of this ARMA model is shown in Figure~\ref{fig:arma_signals}(a), with mean and variance chosen to match that of the original data.  This realization is not totally satisfactory because it is Gaussian without large bursts, and the signal sometimes is negative while the original signal is positive.

A better strategy for this data involves the log transform.  A histogram of the logarithm of the original heat flux is shown in Figure~\ref{fig:turbulent_flux}(c).  This histogram is approximately centered and much closer to normally distributed.  Hence, the logarithm of the heat flux is more appropriate for modeling with a Gaussian process.  Therefore, our strategy will be to estimate the ARMA($p,q$) parameters on the log of the heat flux, generate a signal $\wt{n}(t) = X_t$, and then perform the inverse transform by exponentiation.  The log transform requires that the data be positive.  But this turns out to not be that restrictive an assumption, because if the data is negative, one can add a constant beforehand.  The ARMA parameters estimated from the logarithm of the heat flux are $p=3$, $q=2$, $\vp_1 = 2.1$, $\vp_2 = -1.6$, $\vp_3 = 0.44$, $\th_1 = 0.57$, and $\th_2 = -0.07$.

Given the ARMA-generated signal, we can combine it with an analytic model to represent a noisy flux signal.  The variance $\s^2$ of the generated signal can be scaled to match the variance of the original (log-transformed) data, or increased or decreased as desired.    If the externally given value is $s$, then the naive way to represent the noisy signal is $\wt{s} = s e^{\wt{n}}$.  However, this formula has the undesirable feature of setting $E(\wt{s}) \ge s$, because $E(\wt{n}) \ge 1$ as a result of $\wt{n}$ being centered, the convexity of the exponential, and Jensen's inequality.  More precisely, if $\wt{n} \sim N(0, \s^2)$, then $E(e^{\wt{n}}) = e^{\s^2 / 2}$.  We incorporate this convexity correction and instead use the noisy signal
	\begin{equation}
		\wt{s}(t) = s e^{\wt{n}(t) - \s^2 / 2}.
		\label{arma:convexitycorrection}
	\end{equation}
Then, $\wt{s}$ will be log-normally distributed with an expected value of $s$.
 
A realization of the logarithm-based ARMA model is shown in Figure~\ref{fig:arma_signals}(b).  This model is clearly improved as compared to the ARMA model of the original data in several respects.  Compare with Figures~\ref{fig:turbulent_flux}(a) and \ref{fig:arma_signals}(a).  Because of the logarithm and exponentiation, the generated signal has no negative values, just as in the original signal.  Second, the data has a long tail, with large bursts occurring regularly.  Because using the logarithm-based model is clearly superior to the non-logarithm-based model for generating fluctuations more closely resembling those from the turbulent simulation, and has almost no extra complexity in practice, we will use it for further studies.  This ARMA-based method enables the very efficient generation of arbitrarily long sequences of noise, and will be used heavily in Section \ref{sec:numerical_behavior}.

This method for generating fluctuations to incorporate into an analytic heat flux is useful for some purposes, but cannot model all possible effects.  We have assumed stationarity of the fluctuations, so one behavior not captured is when the properties of the fluctuations, such as their amplitude or time between bursts, depend strongly on the transport solution.  Also, for simplicity, the methods described here involve either finite spatial or temporal correlation, but not both simultaneously.

\section{Behavior of the multiple-timescale coupling method in the presence of fluctuations}
\label{sec:numerical_behavior}
\subsection{Statement of the problem}
We have described both the coupling method and a way to generate fluctuations.  In this section, we combine these elements and apply them to a solvable problem in which the flux is specified analytically.  By so doing, we can characterize the behavior of the coupling method in the presence of  fluctuations, including the convergence of the residual and the error.  The example problem is taken from Ref.~\cite{shestakov:2003}.  Equation~\eqref{paradigm_transport} is to be solved for the steady-state solution, where for this example we choose the flux $q[p]$ to be
	\begin{equation}
		q = - D \pd{p}{x}, \label{flux:shestakov}
	\end{equation}
where
	\begin{equation}
		D = \left( \frac{1}{p} \pd{p}{x} \right)^2.
	\end{equation}
This flux is nonlinear and diffusive.  With a source and boundary conditions of
	\begin{align}
		S(x) &= \begin{cases}
									S_0 & 0 \le x \le a \\
									0 & a < x \le 1
							\end{cases}, \\
		p'(0) &= 0, \\
		p(1) &= P_0,
	\end{align}
the steady-state solution is given by
	\begin{equation}
		p(x) = \begin{cases}
									\left( P_0^{1/3} + \frac{1}{3} (S_0 a)^{1/3} \left[ 1 - a + \frac{3}{4}\left(a -  \frac{x^{4/3} }{a^{1/3}} \right) \right] \right)^3 	& 	0 \le x \le a \\
									\left[ P_0^{1/3} + \frac{1}{3} (S_0 a)^{1/3} (1 - x) \right]^3 		& a < x \le 1
								\end{cases}.
							\label{shestakov:analyticsolution}
	\end{equation}
	
The Neumann boundary condition at $x=0$ reflects the geometric no-flux condition at the magnetic axis of a tokamak.  We use $S_0 = 1$, $a=0.1$, and $P_0 = 0.01$.  The convergence results we present are not that sensitive to the initial condition.  We use a single timestep with $\D t \to \infty$, so we are looking for a steady-state solution, and we therefore suppress the $m$ subscript used in \eref{iteration_scheme}.  The $l$ subscript remains for the Picard-like iteration to converge to the self-consistent solution of \eref{iteration_scheme}.  We take the problem as specified, except now we add fluctuations to the heat flux to examine convergence of the residual and the error.

With $M$ and $g$ defined as in \eref{iteration_scheme_matrix_equation}, we define the residual at iteration $l$ as
	\begin{equation}
		(\text{residual})_l = \operatorname{rms}\left( \frac{M_l p_l - g}{\operatorname{max}(|g|)} \right),
	\end{equation}
where the maximum and rms are taken over the spatial index.  Before defining the numerical error, we observe that if we compared $p_l$ to the analytic solution in \eref{shestakov:analyticsolution}, an additional source of error besides the random noise would be the discretization error.  We use 500 spatial points so that the discretization error here is small, but we wish to eliminate that error entirely from current consideration.  Therefore, we define $p_{\text{exact}}$ as the exact steady-state solution of the discretized version of the transport problem in \eref{paradigm_transport}, solved numerically to high precision.  We compute the error by comparing $p_l$ with $p_{\text{exact}}$ rather than with the analytic solution, so we define
\begin{equation}
		(\text{error})_l = \operatorname{rms} \left( \frac{p_l - p_{\text{exact}}}{\operatorname{max}(p_{\text{exact}})} \right).
	\end{equation}

\subsection{Gaussian noise: spatially correlated, temporally white}
First, we add Gaussian fluctuations in the manner described by \eref{discrete_langevin_combined}.  At each iteration, the spatial dependence of the heat flux is modified as
	\begin{equation}
		\wt{q}(x) = q(x) \bigl[ 1 + W(x) z(x) \bigr].
	\end{equation}
The Gaussian noise $z(x)$ is chosen to have a spatial correlation length of $0.2$.  It is temporally white, so that at each iteration step in the transport solver, a new realization of $z(x)$ is used.  We solve \eref{paradigm_transport} with the modified flux for the steady solution.

\begin{figure}
	\centering
	\includegraphics{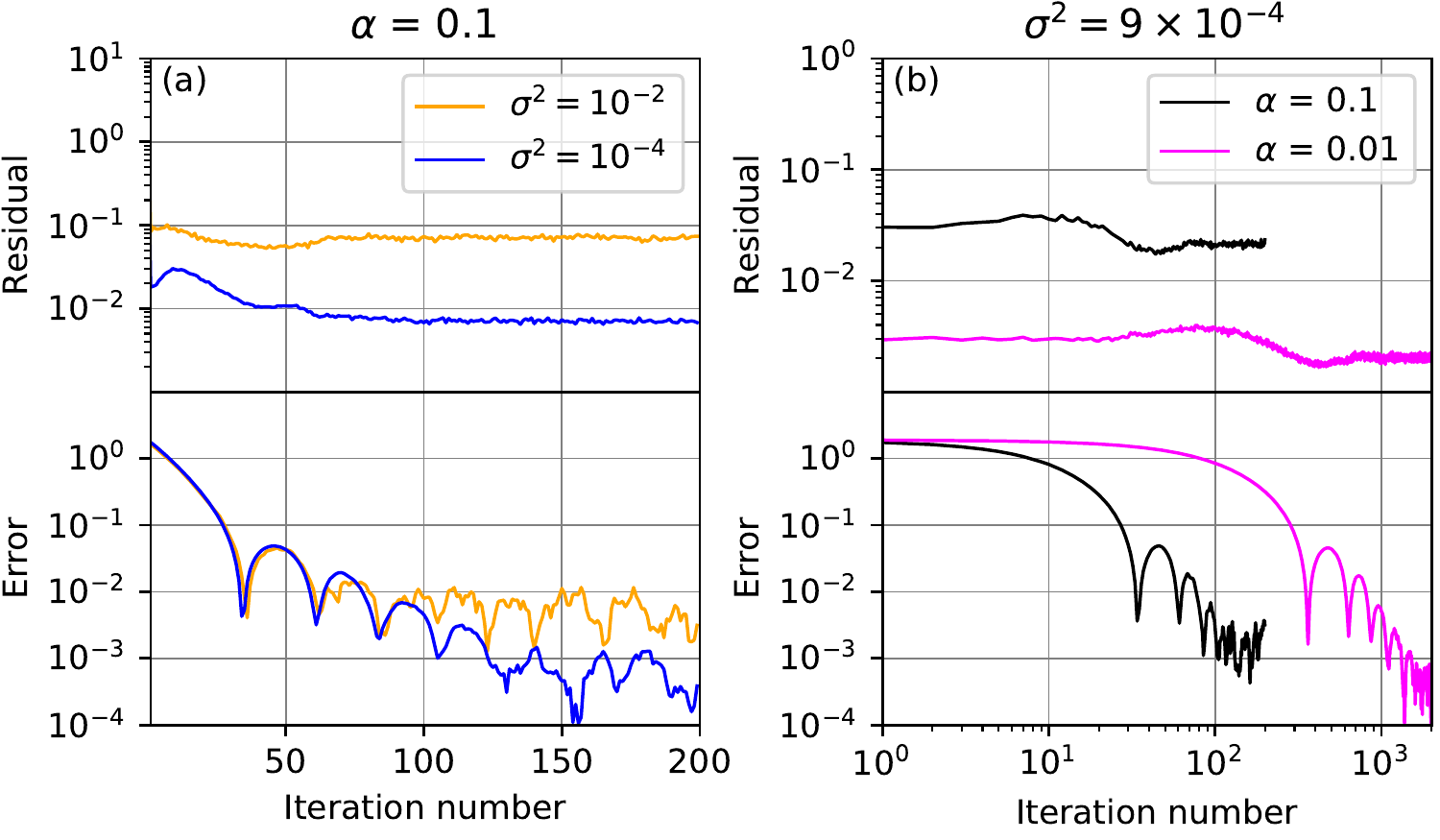}
	\caption{Residual and error of realizations of system with added Gaussian noise, spatially correlated and temporally white.  (\textbf{a}) Two values of variance $\s^2$ of added noise, with fixed relaxation parameter $\a$.  (\textbf{b}) Two values of relaxation parameter, with fixed noise amplitude.  Note that the $x$-axis in (b) uses a logarithmic scale because the system converges in fewer iterations at $\a = 0.1$ than at $\a = 0.01$.}
	\label{fig:tango_gaussian_noise}
\end{figure}

Figure \ref{fig:tango_gaussian_noise} shows the residual and error for this system.  Panel (a) shows two realizations with different amplitudes of the noise $z$ and fixed relaxation parameter $\a$.  At a lower level of noise, the residual and the error both decrease to lower levels, as expected.  Panel (b) shows two realizations with different values of $\a$ and fixed amplitude of the noise.  At the smaller value of $\a$, the residual and error have reached a noise floor at a small level, as expected.

Observe that in all four realizations shown, the residual has reached a noise floor at an earlier iteration number than the error reaches a noise floor.  For example, in panel (a) with noise level of 0.01, the residual flattens around iteration 70, and the error decreases until iteration 150 or so.  This behavior is as predicted by the simple model in Section \ref{sec:iteration_analytic_model} and underscores the point that the behavior of the residual does not necessarily reflect the behavior of the error in the presence of fluctuations.

\subsection{Non-Gaussian noise: temporally correlated, spatially uniform}
For the second method, we use added non-Gaussian, temporally correlated noise.  Given $q(x)$ as in \eref{flux:shestakov}, we first create a time series $q(x,t)$ to mimic the flux from a turbulence simulation.  This time series is created using the method described in Section~\ref{sec:generating_non_gaussian_noise}.  Due to boundary conditions, the noisy flux signal used here involves a slight modification of \eref{arma:convexitycorrection} to apply a windowing function, and is given by
	\begin{equation}
		q(x, t) = q(x) e^{W(x) \wt{n}(t) - W(x)^2 \s^2 /2}.
		\label{tango_noisyflux}
	\end{equation}
$\wt{n}(t)$ is generated using the same procedure and ARMA($p,q$) coefficients as in Figure~\ref{fig:arma_signals}(b), except the variance $\s^2$ of $\wt{n}$ may be scaled from the original value.

We remark on having time-dependent noise $q(x,t)$ while simultaneously looking for a time-independent steady-state solution, $\D t \to \infty$ (although the following considerations apply even for finite $\Delta t$).  Within the context of the multiple-timescale approach, there is no contradiction.  The $t$ in $q(x,t)$ is formally a fast-turbulent-timescale variable, and the $t$ of the transport equation, to which $\D t$ is related, is formally an independent, slow timescale variable.  However, \emph{in practice}, a transport solver must deal with the time dependence of the turbulent heat flux $q(x, t)$.  As alluded to in Section~\ref{sec:iteration_analytic_model}, in coupling turbulence simulations with transport, it is natural for the flux that is passed from the turbulence simulation to the transport solver to be averaged over a time $T$ to yield $\avg{q}$, as in \eref{flux_time_avg}.  We use $\avg{q}$ as the heat flux at each $l$ iteration of Eqs.~\eqref{iteration_scheme}--\eqref{iteration_scheme_pbar}; that is, $q[\ol{p}_l]$ is taken to be $\avg{q}$.  The time dependence in \eref{tango_noisyflux} mimics the time dependence on the turbulent timescale, which is then time-averaged to obtain $\avg{q}$.

The averaging time $T$ is a free parameter.  Given the discussion in Section~\ref{sec:math_considerations} and the central limit theorem, one is motivated to choose an averaging $T$ of at least a few autocorrelation times.  A larger $T$ reduces the variance of $\avg{q}$ and also, because of the central limit theorem, brings its distribution closer to Gaussian.  In our numerical experiments, we will explore the effect of the averaging time $T$ in addition to the amplitude of the fluctuations.  We remark that as long as $T \gg \t$, the correlation between subsequent samples of $\avg{q}$ is negligible.  But this would not be the case for small $T$, and furthermore, the bursty, non-Gaussian behavior remains.



\begin{figure}
	\centering
	\includegraphics{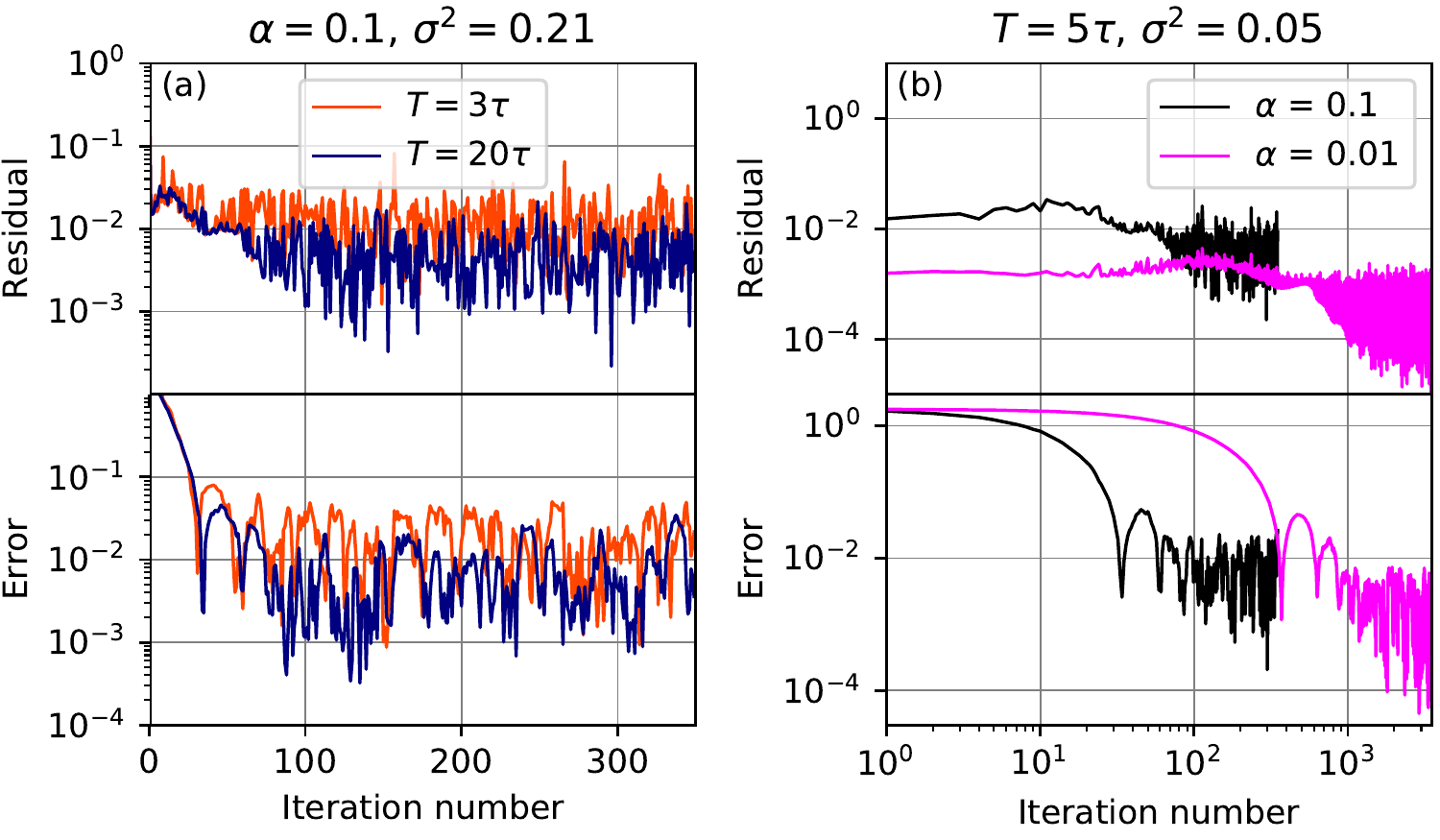}
	\caption{Residual and error of realizations of system with added non-Gaussian noise, spatially uniform and temporally correlated.  $\s^2$ is the variance of the ARMA-generated Gaussian signal $\wt{n}$ in \eref{tango_noisyflux}.  (\textbf{a}) Two values of normalized averaging time $T/\t$, where $\t$ is the autocorrelation time, with fixed relaxation parameter and noise amplitude.  (\textbf{b}) Two values of relaxation parameter, with fixed averaging time and noise amplitude.  Note that the $x$-axis in (b) uses a logarithmic scale because the system converges in fewer iterations at $\a = 0.1$ than $\a = 0.01$.}
	\label{fig:tango_nongaussian_noise}
\end{figure}

Figure~\ref{fig:tango_nongaussian_noise} shows the convergence of the residual and error in realizations of the system with added non-Gaussian noise.  Panel (a) shows two realizations for different averaging times, $T = 3\t$ and $T = 20\t$.  At larger $T$, the error and residual at the noise floor are reduced.  Panel (b) shows two realizations with different values of the relaxation parameter $\a$.  The same behavior as in Figure~\ref{fig:tango_gaussian_noise} is observed, where at smaller values of $\a$, the system takes longer to converge but eventually reaches a smaller residual and error.  We also observe that at $\a = 0.01$, the residual reaches a noise floor at iteration $\approx 700$, while the error continue to decrease until it reaches a noise floor at iteration $\approx 1200$.  This behavior where the error can continue to decrease even when the residual does not is consistent with the simple model in Section~\ref{sec:iteration_analytic_model}.

\begin{figure}
	\centering
	\includegraphics{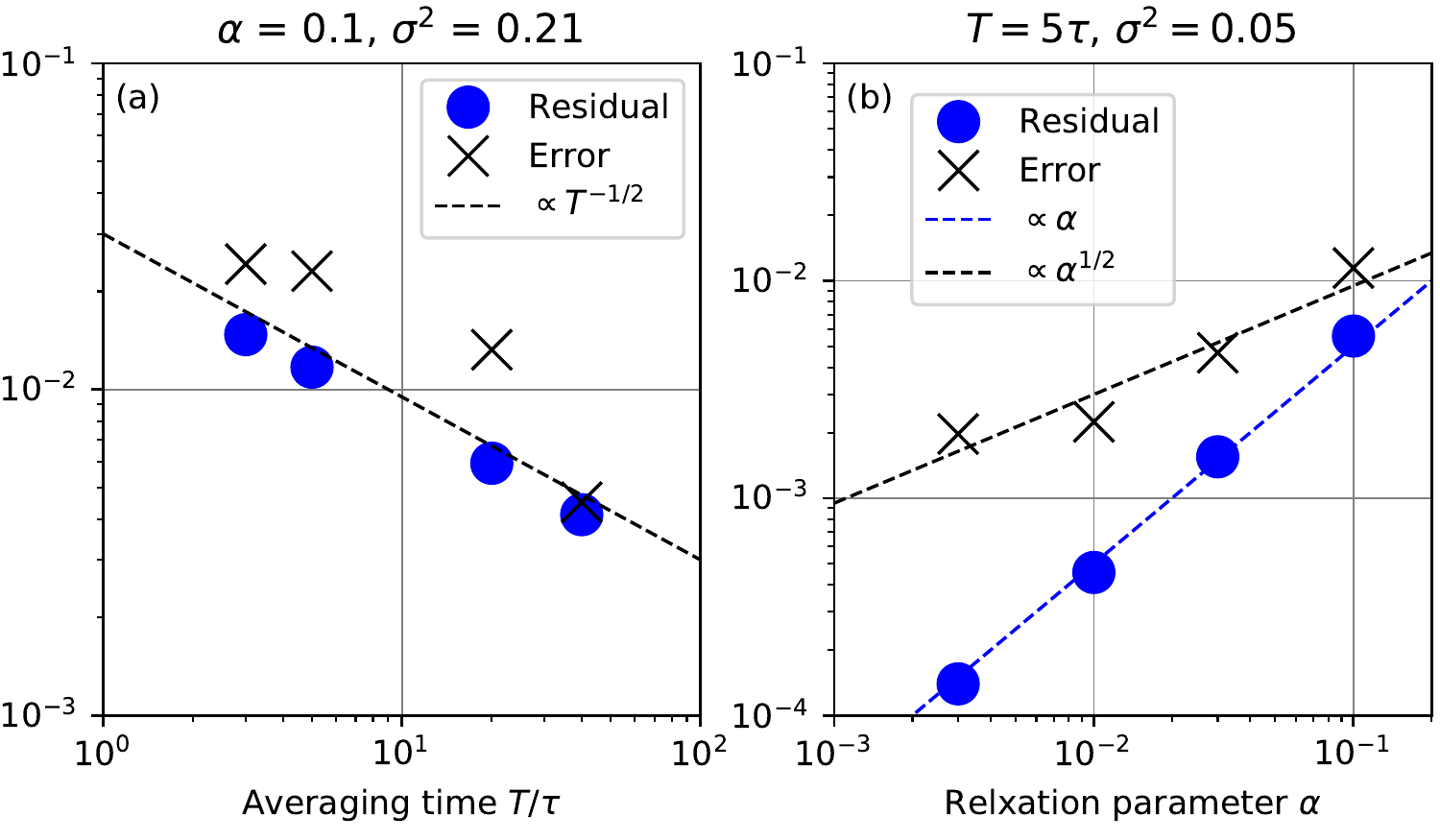}
	\caption{Residual and error at the noise floor.  (\textbf{a}) Scaling with normalized averaging time $T/\t$, where $\t$ is the autocorrelation time.  Both the residual and the error scale as $\sim T^{-1/2}$.  This scaling is consistent with the simple model in Section~\ref{sec:iteration_analytic_model} because the standard deviation of the averaged noise is proportional to $T^{-1/2}$.  (\textbf{b}) Scaling with the relaxation parameter $\a$.  The error scales as $\sim \a^{1/2}$ and the residual scales as $\sim \a$.  The two different scalings is consistent with the simple analytic model in Section~\ref{sec:iteration_analytic_model}.}
	\label{fig:different_error_scalings}
\end{figure}

In Figure~\ref{fig:different_error_scalings}, we show the residual and the error at the noise floor and their scaling with the averaging time and the relaxation parameter.  The residual and the error at the noise floor scale as $T^{-1/2}$, which is expected because the standard deviation of the fluctuations also scales as $T^{-1/2}$.  As the relaxation parameter varies, the residual is proportional to $\a$ while the error is proportional to $\a^{1/2}$.  Again, this behavior is as predicted by the model in Section~\ref{sec:iteration_analytic_model}.  For these plots, we have averaged the error of each iteration over the final 10\% of iterations, after the noise floor has been reached.  This definition gives a measure of the instantaneous error and allows us to make contact with the scaling found from our earlier analytic model.  In practice, averaging the solution over many iterations will provide a better answer, with a smaller error.

\begin{figure}
	\centering
	\includegraphics{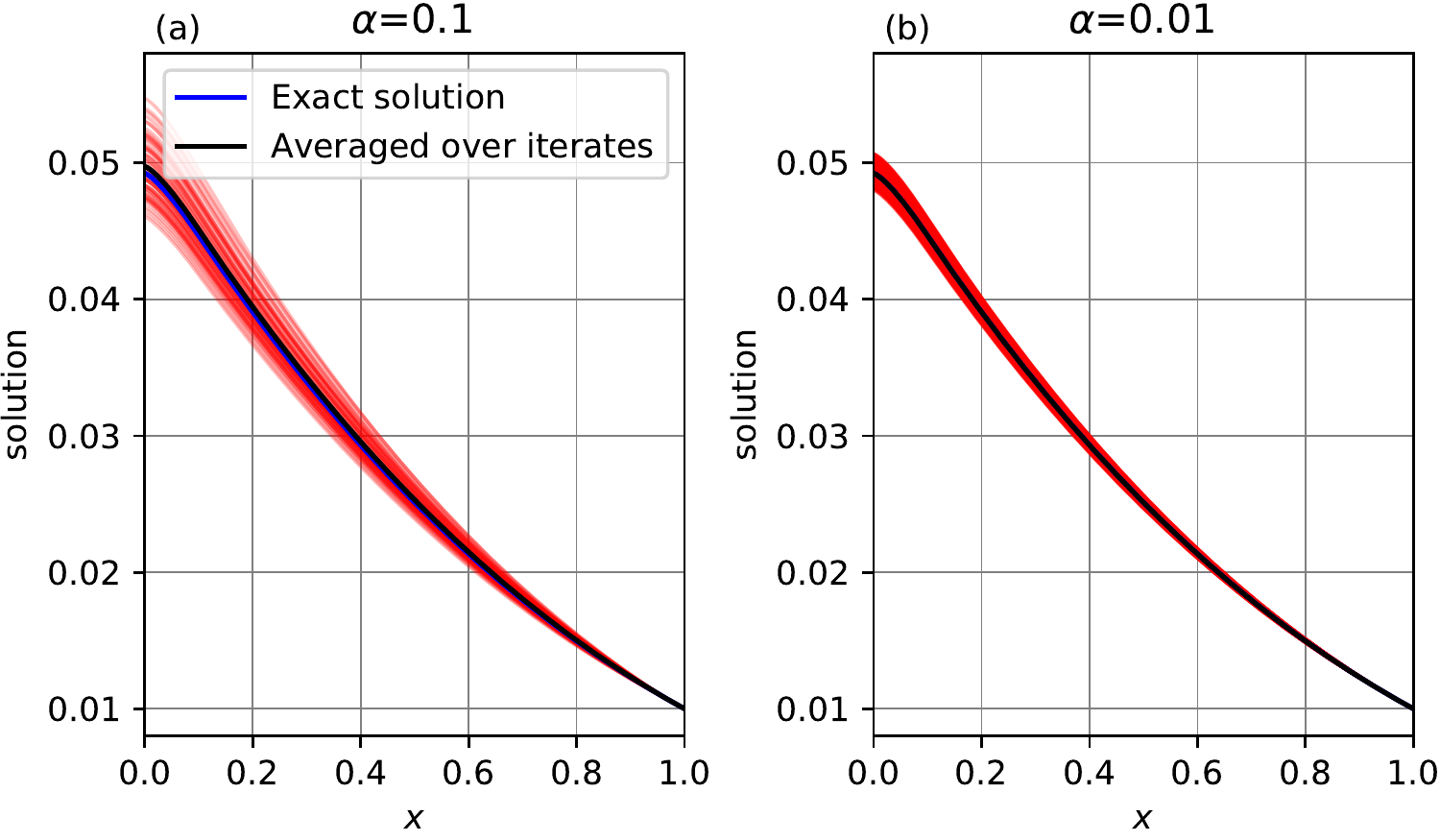}
	\caption{At $T=5\t$ and $\s^2 = 0.21$.  (\textbf{a}) Relaxation parameter $\a = 0.1$, showing the profile in the final 60 iterates (red lines), the average profile over those 60 iterates (black line), and the exact solution (blue lines).  The exact solution and averaged profile are nearly visually overlapping.  (\textbf{b}) Same as (a) but with $\a = 0.01$ and showing 600 iterates.  Here, the averaged profile and the exact solution are visually indistinguishable.  In each iterate, the error is zero at $x=1$ where a Dirichlet boundary condition is applied, and largest at $x=0$, where a Neumann boundary condition is applied.}
	\label{fig:profile_distribution}
\end{figure}

Figure~\ref{fig:profile_distribution} presents an illustration of this averaging, showing the profile at several individual iterates (red lines), the profile averaged over those iterates (black line), and the exact solution (blue line).  For $\a=0.1$, 60 iterates are shown, and the averaged profile is almost indistinguishable from the exact solution.  For $\a=0.01$, 600 iterates are shown, which have a narrower spread than at $\a=0.1$, and the averaged profile is visually indistinguishable from the exact solution.

\section{Discussion}
\label{sec:discussion}
We have explored the effect of fluctuations on convergence of a coupled turbulence-transport simulation.  Our motivation is the transport caused by gyrokinetic turbulence in the core of a tokamak.  The fluctuations inherent in direct numerical simulations (DNS) of turbulence present a challenge for convergence of transport solvers, and the error induced by the fluctuations must be understood and managed.  To study this problem, we set up a simpler model of a transport solver with flux computed not by a turbulence simulation, but by an analytic formula \emph{plus} artificially generated noise.

In treating turbulent fluctuations within a statistical framework, we have identified the autocorrelation time $\t$ of the fluctuations as a key parameter.  A coupled turbulence-transport simulation will pass the turbulent flux, averaged over a time $T$, to the transport solver.  $T$ is naturally measured in terms of $\t$.  Furthermore, the central limit theorem for correlated sequences shows that in a stationary process, the averaged flux becomes normally distributed as $T \to \infty$.

To investigate the effect of fluctuations on convergence without the computational expense of full turbulence DNS, we have proposed a method for generating random noise to be combined with prescribed fluxes.  We have used classic autoregressive-moving-average (ARMA) modeling from time-series analysis, which is simple, well understood, and easily implemented with available software packages.  With appropriate transforms, non-Gaussian signals can be represented.

Adding these fluctuations to a solvable transport problem enables us to examine the behavior of a numerical method for solving the coupled turbulence-transport system.  Using either spatially correlated or temporally correlated noise, we have explored the convergence of the residual and the error as we vary the amplitude of the noise, $T$, and the relaxation parameter $\a$.  One conclusion is that when relaxation is used, the residual may not be a good proxy for the error of the solution.  The residual can reach a noise floor when further iterations decrease the error.  Moreover, at the noise floor, the amplitude of the residual scales as $\a$, whereas the amplitude of the error scales as $\a^{1/2}$.  The behaviors can be understood through a simple, one-degree-of-freedom analytic model.

For actual turbulence-transport simulations, both $\a$ and $T$ are control knobs.  Smaller $\a$ and larger $T$ provide more effective averaging, and the error at the noise floor scales as $\a^{1/2}$ and $T^{-1/2}$.  On the other hand, the total computational time to converge scales as $\a^{-1} T$.  Hence, in terms of scaling alone, neither parameter is a clearly superior control knob.  In practice, $\a$ will have to be small enough to stabilize the iterations \cite{shestakov:2003}.

This sort of testing with artificial fluctuations allows us to predict approximately how much averaging is required to achieve a given accuracy in the real problem of transport-turbulence simulations.  Assuming the fluctuations in DNS remain comparable to those in the ARMA training data, our results here suggest that for $T = 5\t$ and $\a = 0.1$, the error at the noise floor is about a few percent.  The analogous prediction for $\a=0.3$ was borne out in the coupled turbulence-transport simulations in Ref.~\cite{parker:2018}.  Hence, these approximate predictions with simple models can provide a very useful guide.  On the other hand, gyrokinetic simulations in some situations, such as close to marginal stability, have been known to be bursty and intermittent, with long periods of quiescence followed by short-duration bursts that dominate the total flux.  Such a regime, consisting of fluctuations qualitatively different from those used in this study, would be challenging for any multiple-timescale method.  It remains to be seen how well the method here would fare in that regime.

In summary, one challenge of coupling turbulence simulations directly to a transport solver is how fluctuations in the simulated turbulent flux affect convergence.  At least some of the effect of the fluctuations can be understood through relatively simple models and artificially generated noise, as we have done here.  For characterizing the behavior of a transport solver in the presence of fluctuations, such simple models provide an efficient alternative to using computationally intensive turbulence simulations.



\vspace{6pt} 



\funding{This work was performed under the auspices of the U.S.\ Department of Energy by Lawrence Livermore National Laboratory under Contract No.\ DE-AC52-07NA27344, in part through the SciDAC Partnership for Multiscale Gyrokinetic Turbulence.  This research was supported by the Exascale Computing Project (17-SC-20-SC), a collaborative effort of the U.S. Department of Energy Office of Science and the National Nuclear Security Administration.}

\acknowledgments{We acknowledge useful discussions with Lee Ricketson and Gabriele Merlo.}

\conflictsofinterest{The authors declare no conflict of interest.  The founding sponsors had no role in the design of the study; in the collection, analyses, or interpretation of data; in the writing of the manuscript, and in the decision to publish the results.} 



\appendixtitles{yes} 
\appendixsections{one} 

\appendix
\section{Estimating the autocorrelation time}
Estimating the autocorrelation time from a sample is less trivial than estimating the mean.  Reference~\cite{thompson:2010} discusses several methods for estimating $\t$; two are reviewed here.

One option is the \emph{batch means} method.  In the batch means method, a signal is split into batches of size $m$.  For large $m$, the set of means of the batches has asymptotic variance $\s^2 \t / m$.  If $s^2$ is the sample variance of all of the $X_i$, and $s_m^2$ is the sample variance of the batch means, then the autocorrelation time can be estimated as
	\begin{equation}
		\t \approx \frac{m s_m^2}{s^2}.
	\end{equation}
For a signal with $n$ samples, Ref.~\cite{thompson:2010} proposed using $n^{1/3}$ batches of size $m=n^{2/3}$.  

Another method is to first fit an ARMA model to the data, as described in Section~\ref{sec:generating_non_gaussian_noise}, then to compute the autocorrelation time of the corresponding ARMA process.  Reference~\cite{thompson:2010} gives a formula for $\t$ for an AR($p$) process.  For a more general ARMA($p,q$) model, one can calculate \cite{wei:book}
	\begin{equation}
		\t = (1 - \vp_1 \r_1 - \cdots - \vp_p \r_p) \frac{(1 - \th_1 - \cdots - \th_q)^2}{1 - \vp_1 - \cdots - \vp_p)^2},
	\end{equation}
where $\vp_j$ and $\th_j$ are the AR and MA parameters, and $\r_j$ are the autocorrelation coefficients of the ARMA model.


\reftitle{References}

\externalbibliography{yes}
\bibliography{multiscale_methods}

\begin{thebibliography}{-------}
\providecommand{\natexlab}[1]{#1}

\bibitem[Kim \em{et~al.}(2016)Kim, Bulmer, Campbell, Casper, LoDestro, Meyer,
  Pearlstein, and Snipes]{kim:2016}
Kim, S.; Bulmer, R.; Campbell, D.; Casper, T.; LoDestro, L.; Meyer, W.;
  Pearlstein, L.; Snipes, J.
\newblock CORSICA modelling of ITER hybrid operation scenarios.
\newblock {\em Nucl. Fusion} {\bf 2016}, {\em 56},~126002.

\bibitem[Tang(1986)]{tang:1986}
Tang, W.
\newblock Microinstability-based model for anomalous thermal confinement in
  tokamaks.
\newblock {\em Nucl. Fusion} {\bf 1986}, {\em 26},~1605.

\bibitem[Jardin \em{et~al.}(1993)Jardin, Bell, and Pomphrey]{jardin:1993}
Jardin, S.; Bell, M.; Pomphrey, N.
\newblock TSC simulation of Ohmic discharges in TFTR.
\newblock {\em Nucl. Fusion} {\bf 1993}, {\em 33},~371.

\bibitem[Erba \em{et~al.}(1997)Erba, Cherubini, Parail, Springmann, and
  Taroni]{erba:1997}
Erba, M.; Cherubini, A.; Parail, V.V.; Springmann, E.; Taroni, A.
\newblock Development of a non-local model for tokamak heat transport in
  L-mode, H-mode and transient regimes.
\newblock {\em Plasma Phys. Control. Fusion} {\bf 1997}, {\em 39},~261.

\bibitem[Kinsey \em{et~al.}(2011)Kinsey, Staebler, Candy, Waltz, and
  Budny]{kinsey:2011}
Kinsey, J.; Staebler, G.; Candy, J.; Waltz, R.; Budny, R.
\newblock ITER predictions using the GYRO verified and experimentally validated
  trapped gyro-Landau fluid transport model.
\newblock {\em Nucl. Fusion} {\bf 2011}, {\em 51},~083001.

\bibitem[Bourdelle \em{et~al.}(2007)Bourdelle, Garbet, Imbeaux, Casati, Dubuit,
  Guirlet, and Parisot]{bourdelle:2007}
Bourdelle, C.; Garbet, X.; Imbeaux, F.; Casati, A.; Dubuit, N.; Guirlet, R.;
  Parisot, T.
\newblock A new gyrokinetic quasilinear transport model applied to particle
  transport in tokamak plasmas.
\newblock {\em Phys. Plasmas} {\bf 2007}, {\em 14},~112501,
  \href{http://xxx.lanl.gov/abs/https://doi.org/10.1063/1.2800869}{{\normalfont
  [https://doi.org/10.1063/1.2800869]}}.
\newblock
  doi:{\changeurlcolor{black}\href{https://doi.org/10.1063/1.2800869}{\detokenize{10.1063/1.2800869}}}.

\bibitem[Citrin \em{et~al.}(2015)Citrin, Breton, Felici, Imbeaux, Aniel,
  Artaud, Baiocchi, Bourdelle, Camenen, and Garcia]{citrin:2015}
Citrin, J.; Breton, S.; Felici, F.; Imbeaux, F.; Aniel, T.; Artaud, J.;
  Baiocchi, B.; Bourdelle, C.; Camenen, Y.; Garcia, J.
\newblock Real-time capable first principle based modelling of tokamak
  turbulent transport.
\newblock {\em Nucl. Fusion} {\bf 2015}, {\em 55},~092001.

\bibitem[Meneghini \em{et~al.}(2017)Meneghini, Smith, Snyder, Staebler, Candy,
  Belli, Lao, Kostuk, Luce, Luda, Park, and Poli]{meneghini:2017}
Meneghini, O.; Smith, S.; Snyder, P.; Staebler, G.; Candy, J.; Belli, E.; Lao,
  L.; Kostuk, M.; Luce, T.; Luda, T.; Park, J.; Poli, F.
\newblock Self-consistent core-pedestal transport simulations with neural
  network accelerated models.
\newblock {\em Nucl. Fusion} {\bf 2017}, {\em 57},~086034.

\bibitem[Shestakov \em{et~al.}(2003)Shestakov, Cohen, Crotinger, LoDestro,
  Tarditi, and Xu]{shestakov:2003}
Shestakov, A.; Cohen, R.; Crotinger, J.; LoDestro, L.; Tarditi, A.; Xu, X.
\newblock Self-consistent modeling of turbulence and transport.
\newblock {\em J. Comput. Phys.} {\bf 2003}, {\em 185},~399 -- 426.
\newblock
  doi:{\changeurlcolor{black}\href{https://doi.org/http://dx.doi.org/10.1016/S0021-9991(02)00063-3}{\detokenize{http://dx.doi.org/10.1016/S0021-9991(02)00063-3}}}.

\bibitem[Candy \em{et~al.}(2009)Candy, Holland, Waltz, Fahey, and
  Belli]{candy:2009}
Candy, J.; Holland, C.; Waltz, R.E.; Fahey, M.R.; Belli, E.
\newblock Tokamak profile prediction using direct gyrokinetic and neoclassical
  simulation.
\newblock {\em Phys. Plasmas} {\bf 2009}, {\em 16},~060704.
\newblock
  doi:{\changeurlcolor{black}\href{https://doi.org/10.1063/1.3167820}{\detokenize{10.1063/1.3167820}}}.

\bibitem[Barnes \em{et~al.}(2010)Barnes, Abel, Dorland, G\"{o}rler, Hammett,
  and Jenko]{barnes:2010}
Barnes, M.; Abel, I.G.; Dorland, W.; G\"{o}rler, T.; Hammett, G.W.; Jenko, F.
\newblock Direct multiscale coupling of a transport code to gyrokinetic
  turbulence codes.
\newblock {\em Phys. Plasmas} {\bf 2010}, {\em 17},~056109.
\newblock
  doi:{\changeurlcolor{black}\href{https://doi.org/10.1063/1.3323082}{\detokenize{10.1063/1.3323082}}}.

\bibitem[Parker \em{et~al.}(2018)Parker, LoDestro, Told, Merlo, Ricketson,
  Campos, Jenko, and Hittinger]{parker:2018}
Parker, J.B.; LoDestro, L.L.; Told, D.; Merlo, G.; Ricketson, L.F.; Campos, A.;
  Jenko, F.; Hittinger, J.A.
\newblock Bringing global gyrokinetic turbulence simulations to the transport
  timescale using a multiscale approach.
\newblock {\em Nucl. Fusion} {\bf 2018}, {\em 58},~054004.

\bibitem[Highcock \em{et~al.}(2018)Highcock, Mandell, Barnes, and
  Dorland]{highcock:2018}
Highcock, E.G.; Mandell, N.R.; Barnes, M.; Dorland, W.
\newblock Optimisation of confinement in a fusion reactor using a nonlinear
  turbulence model.
\newblock {\em J. Plasma Phys.} {\bf 2018}, {\em 84},~905840208.
\newblock
  doi:{\changeurlcolor{black}\href{https://doi.org/10.1017/S002237781800034X}{\detokenize{10.1017/S002237781800034X}}}.

\bibitem[Sugama and Horton(1997)]{sugama:1997}
Sugama, H.; Horton, W.
\newblock Transport processes and entropy production in toroidally rotating
  plasmas with electrostatic turbulence.
\newblock {\em Phys. Plasmas} {\bf 1997}, {\em 4},~405--418.
\newblock
  doi:{\changeurlcolor{black}\href{https://doi.org/10.1063/1.872099}{\detokenize{10.1063/1.872099}}}.

\bibitem[Sugama and Horton(1998)]{sugama:1998}
Sugama, H.; Horton, W.
\newblock Nonlinear electromagnetic gyrokinetic equation for plasmas with large
  mean flows.
\newblock {\em Phys. Plasmas} {\bf 1998}, {\em 5},~2560--2573.
\newblock
  doi:{\changeurlcolor{black}\href{https://doi.org/10.1063/1.872941}{\detokenize{10.1063/1.872941}}}.

\bibitem[Abel \em{et~al.}(2013)Abel, Plunk, Wang, Barnes, Cowley, Dorland, and
  Schekochihin]{abel:2013}
Abel, I.G.; Plunk, G.G.; Wang, E.; Barnes, M.; Cowley, S.C.; Dorland, W.;
  Schekochihin, A.A.
\newblock Multiscale gyrokinetics for rotating tokamak plasmas: fluctuations,
  transport and energy flows.
\newblock {\em Rep. Prog. Phys.} {\bf 2013}, {\em 76},~116201.

\bibitem[tan()]{tangocode}
 \href{http://xxx.lanl.gov/abs/https://github.com/LLNL/tango}{{\normalfont
  [https://github.com/LLNL/tango]}}.

\bibitem[Billingsley(1995)]{billingsley:book}
Billingsley, P.
\newblock {\em Probability and Measure}; Wiley Series in Probability and
  Statistics, Wiley,  1995.

\bibitem[Jenko \em{et~al.}(2000)Jenko, Dorland, Kotschenreuther, and
  Rogers]{jenko:2000}
Jenko, F.; Dorland, W.; Kotschenreuther, M.; Rogers, B.N.
\newblock Electron temperature gradient driven turbulence.
\newblock {\em Phys. Plasmas} {\bf 2000}, {\em 7},~1904--1910.
\newblock
  doi:{\changeurlcolor{black}\href{https://doi.org/10.1063/1.874014}{\detokenize{10.1063/1.874014}}}.

\bibitem[G\"{o}rler \em{et~al.}(2011{\natexlab{a}})G\"{o}rler, Lapillonne,
  Brunner, Dannert, Jenko, Merz, and Told]{gorler:2011}
G\"{o}rler, T.; Lapillonne, X.; Brunner, S.; Dannert, T.; Jenko, F.; Merz, F.;
  Told, D.
\newblock The global version of the gyrokinetic turbulence code GENE.
\newblock {\em J. Comput. Phys.} {\bf 2011}, {\em 230},~7053 -- 7071.
\newblock
  doi:{\changeurlcolor{black}\href{https://doi.org/http://dx.doi.org/10.1016/j.jcp.2011.05.034}{\detokenize{http://dx.doi.org/10.1016/j.jcp.2011.05.034}}}.

\bibitem[G\"{o}rler \em{et~al.}(2011{\natexlab{b}})G\"{o}rler, Lapillonne,
  Brunner, Dannert, Jenko, Aghdam, Marcus, McMillan, Merz, Sauter, Told, and
  Villard]{gorler:2011b}
G\"{o}rler, T.; Lapillonne, X.; Brunner, S.; Dannert, T.; Jenko, F.; Aghdam,
  S.K.; Marcus, P.; McMillan, B.F.; Merz, F.; Sauter, O.; Told, D.; Villard, L.
\newblock Flux- and gradient-driven global gyrokinetic simulation of tokamak
  turbulence.
\newblock {\em Phys. Plasmas} {\bf 2011}, {\em 18},~056103.
\newblock
  doi:{\changeurlcolor{black}\href{https://doi.org/10.1063/1.3567484}{\detokenize{10.1063/1.3567484}}}.

\bibitem[gen()]{genecode}
 \href{http://xxx.lanl.gov/abs/www.genecode.org}{{\normalfont
  [www.genecode.org]}}.

\bibitem[Box \em{et~al.}(2015)Box, Jenkins, Reinsel, and Ljung]{box:book}
Box, G.E.; Jenkins, G.M.; Reinsel, G.C.; Ljung, G.M.
\newblock {\em Time series analysis: forecasting and control}; John Wiley \&
  Sons,  2015.

\bibitem[Wei(2006)]{wei:book}
Wei, W.W.
\newblock {\em Time series analysis}; Addison-Wesley,  2006.

\bibitem[Broersen(2002)]{broersen:2002}
Broersen, P.M.T.
\newblock Automatic spectral analysis with time series models.
\newblock {\em IEEE Trans. Instrum. Meas.} {\bf 2002}, {\em 51},~211--216.

\bibitem[Broersen(2003)]{broersen:2003}
Broersen, P.M.
\newblock Automatic Time Series Identification Spectral Analysis with MATLAB
  Toolbox ARMASA.
\newblock {\em IFAC Proceedings Volumes} {\bf 2003}, {\em 36},~1435 -- 1440.
\newblock 13th IFAC Symposium on System Identification (SYSID 2003), Rotterdam,
  The Netherlands, 27-29 August, 2003,
  doi:{\changeurlcolor{black}\href{https://doi.org/https://doi.org/10.1016/S1474-6670(17)34962-5}{\detokenize{https://doi.org/10.1016/S1474-6670(17)34962-5}}}.

\bibitem[Broersen(Retrieved October 22, 2015)]{broersen:mathworks}
Broersen, P.M.T.
\newblock ARMASA.
\newblock https://www.mathworks.com/matlabcentral/fileexchange/1330-armasa,
  Retrieved October 22, 2015.

\bibitem[Thompson(2010)]{thompson:2010}
Thompson, M.B.
\newblock A Comparison of Methods for Computing Autocorrelation Time.
\newblock {\em arXiv} {\bf 2010},
  \href{http://xxx.lanl.gov/abs/1011.0175}{{\normalfont
  [arXiv:stat.CO/1011.0175]}}.

\end{thebibliography}


\end{document}